\documentclass[showpacs,preprintnumbers,amsmath,amssymb,pra]{revtex4}

\usepackage{amsmath}
\usepackage{amssymb}
\usepackage[dvips]{graphicx}
\usepackage[dvips]{color}
\usepackage[T2A]{fontenc}
\usepackage[cp1251]{inputenc}
\usepackage[english]{babel}

\newcommand{\bra}[1]{\left\langle#1\right\vert}
\newcommand{\ket}[1]{\left\vert#1\right\rangle}

\begin{document}

\title{On Preparing Entangled Pairs of Polarization Qubits in the Frequency Non-Degenerate Regime}
  \author{S.S.Straupe, S.P.Kulik}
  \affiliation{Faculty of Physics M.V.Lomonosov Moscow State University, Moscow, 119992, Vorobjevy Gory 1}

\begin{abstract}

The problems associated with practical implementation of the scheme proposed for preparation of arbitrary states of polarization ququarts based on biphotons are discussed. The influence of frequency dispersion effects are considered, and the necessity of group velocities dispersion compensation in the frequency non-degenerate case even for continuous pumping is demonstrated. A method for this compensation is proposed and implemented experimentally. Physical restrictions on the quality of prepared two-photon states are revealed.

\end{abstract}
\maketitle

\section{Introduction}

So-called high-dimensional quantum systems
or systems with a dimension of Hilbert space $D\geq 3$ have attracted increasing interest. Primarily, this interest stems from the possibility of using such systems in various quantum information protocols. The application of these systems, which are frequently referred to as \emph{qudits} (quantum dits), as information carriers offers a number of advantages over encoding with qubits \cite{PeresPRL, TittelPRA, BjorkPRA, KilinOptSpec}. The second important field of application of qudits is verification of basic principles of quantum theory, in particular, violation of Bell-type inequalities \cite{Bell}. Incidentally, it turns out that the use of high-dimensional systems in a number of cases makes it possible to increase the quantitative gap that arises in the description of composite systems using the language of quantum or classical correlations. In turn, this allows one to reduce the requirements imposed on quantum efficiency of photo-detectors and other experimental equipment \cite{Genovese}. Thus, the development of methods used for preparing and characterizing different states of qudits is an important problem giving rise to a new field of quantum information, i.e., so-called \emph{quantum state engineering}.

Polarization degrees of freedom of photons are especially convenient tools for experimental preparation of qudit states, primarily owing to simplicity of performing polarization transformations. Specifically, in order to control the polarization state, it is sufficient
to use linear optical elements, like phase plates and beam-splitters. Moreover, the developed statistical methods of polarization tomography have made it possible to completely reconstruct the state vector of the initial quantum system as a result of the relatively simple procedure \cite{Kwiat, BogdanovPRA06}.

A polarization ququart is defined to be a pure two-photon state that can be written in the form
\begin{equation}\label{ququart}
    \ket{\psi}= c_1\ket{H_1}\ket{H_2}+c_2\ket{H_1}\ket{V_2}+c_3\ket{V_1}\ket{H_2}+c_4\ket{V_1}\ket{V_2},
\end{equation}
where the ket vectors $\ket{H}$ and $\ket{V}$ stand for the basis single-photon polarization states of the field mode,
and indices $1$ and $2$ can refer to either frequency (longitudinal) or spatial (transverse) modes. It is
common practice to distinguish spatial polarization ququarts, for which the field state is quasi-degenerate
in frequency but there are two spatial modes (see, for example, \cite{Kwiat}), and frequency polarization ququarts,
when the spatial modes are degenerate but the frequency spectrum of the field contains two components
\cite{freqququart}. The first experiments on the preparation of entangled pairs of polarization qubits or entangled states of
ququarts were performed in the mid-90s. In \cite{KiessPRL93, ShihPRA94} pairs of photons generated in collinear
frequency-degenerate spontaneous parametric downconversion (SPDC) with type-II phase-matching were transformed into polarization-momentum entangled states with a beam splitter and subsequent postselection. Shortly thereafter, Kwiat et al. \cite{KwiatPRL95} proposed a method that does not require postselection and is based on the type-II non-collinear phase matching. There are well-known schemes employed for preparation of entangled states with the use of two crystals with type-I phase-matching in non-collinear frequency degenerate \cite{KwiatPRA99} and collinear but frequency non-degenerate \cite{BurlakovJETP02, KulikPRL06} regimes. The case of pulsed pump is discussed in \cite{Kim1, Kim2}. One of the advantages of these schemes is the possibility of preparing biphotons in non-maximally entangled states \cite{KwiatPRL99}. Of special interest, however, are the methods providing
the preparation of polarization ququarts in an \emph{arbitrary} state by means of unitary (lossless) transformations, when all components of the polarization state of a biphoton pair can be completely controlled. The design of this universal source of ququarts is of both fundamental and practical interest for solving problems of quantum information and quantum communication. The experimental scheme that ensures complete control of a polarization ququart was proposed in a recent paper \cite{KimPRA08}. Being essentially an interferometric scheme it provides the stability of parameters, which is not inherent in conventional interferometric schemes but is well
known in classical polarization optics. To the best of our knowledge, this scheme has not been fully implemented so far. However, specific families of states of frequency non-degenerate ququarts, of course, have already been prepared experimentally (see, for example, \cite{BogdanovPRA06, KulikPRL06, Shurupov}).

In this work, we study physical restrictions imposed on the quality of states prepared according to the aforementioned scheme
and discuss the methods proposed for their elimination. It has turned out that the use of the frequency non-degenerate regime and, correspondingly, the polarization and frequency entanglement in the scheme with two consequent crystals involves
specific experimental difficulties due to the frequency dispersion in crystals. However, this scheme is very convenient from the viewpoint of applications in quantum information protocols and has been frequently employed in their implementation. In this paper, we
demonstrate that, in order to achieve high quality of prepared states, it is necessary to use compensators for group velocity dispersion, which were earlier treated as an attribute of schemes with type-II phase matching \cite{KiessPRL93, ShihPRA94, Rubin94} or pulse pumping \cite{Kim1, Kim2} only.

\section{Scheme for preparation of arbitrary polarization states of ququarts}

Let us consider the scheme proposed in \cite{KimPRA08} for the preparation of arbitrary states of polarization ququarts. We start with the formal mathematical aspect of the problem. It is known that an arbitrary pure entangled state of a qubit pair or a ququart can be
represented in the form (cf. expression (\ref{ququart}))
\begin{equation}\label{Schmidt}
    \ket{\psi}= \sqrt{\mu}\ket{A_1}\ket{A_2}+\sqrt{1-\mu}\ket{B_1}\ket{B_2},
\end{equation}
where $\ket{A_j}$ and $\ket{B_j}$ are the basis vectors in single-qubit state spaces. This expression is known as the Schmidt decomposition. The coefficients $\mu, (1-\mu)$ are eigenvalues of the single-particle density matrices of each qubit (which, as is known, coincide with each other), and the vectors $\ket{A_j}$ and $\ket{B_j}$ form an orthogonal basis in which they are diagonal.
Therefore, in order to prepare an arbitrary ququart state, one should be able to experimentally control the coefficients in the Schmidt decomposition and to perform switching between the bases. Note that the former operation changes the degree of entanglement
of the state and affects the degrees of freedom of both qubits. In this sense, such an operation can be termed “nonlocal”\footnote{The term “(non)local” is commonly accepted in the modern
theory of quantum information and, in essence, does not indicate the (presence) absence of instantaneous interactions between subsystems.}. However, the transformation of the basis vectors requires only “local” operations, i.e., operations that act on each qubit separately. It should also be noted that the degree of entanglement of a qubit pair can be defined in any reasonable manner;
for example, it is convenient to use concurrence \cite{Wootters98}.

\begin{figure}[h]
\centering\includegraphics[width=\textwidth]{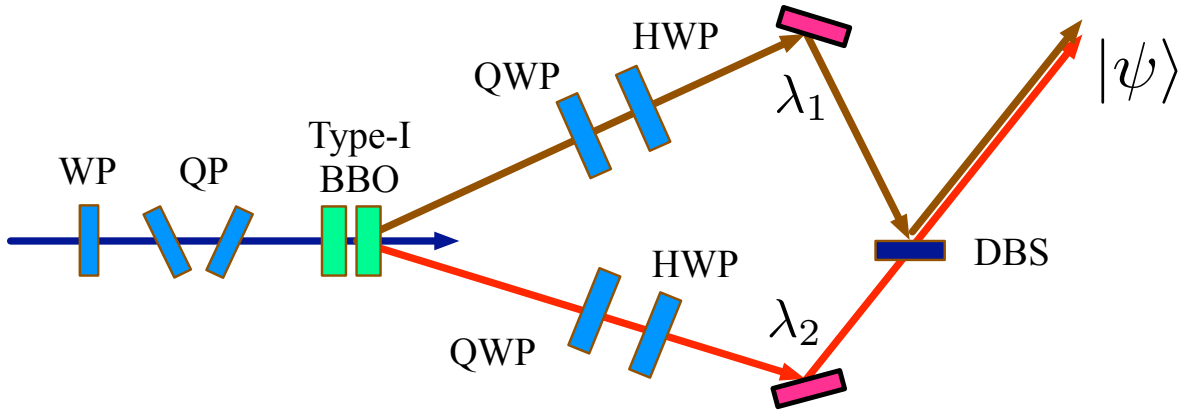}
\caption{{\protect\footnotesize {A scheme for preparation of an arbitrary pure ququart state using two type-I crystals.}}}
\label{proposal2}
\end{figure}

The idea of possible implementation of the scheme used for the preparation of arbitrary states of polarization ququarts is illustrated in Fig. \ref{proposal2}.
This scheme employs two nonlinear crystals cut for non-collinear frequency non-degenerate
type-I phase-matching when both (signal and idler) photons are equally polarized. CW laser is used as a pump, the direction
of linear polarization of the pump radiation is specified using a half-wave plate (WP), and the relative phase between horizontal and vertical components is controlled by tilting a pair of quartz plates (QPs). The state of biphotons generated in the process of spontaneous parametric down-conversion under these conditions has the form
\begin{equation}\label{seed}
    \ket{\psi}= \sqrt{\mu}\ket{H_1}\ket{H_2}+\sqrt{1-\mu}\ket{V_1}\ket{V_2},
\end{equation}
where indices $1$ and $2$ correspond to different frequency modes. The state of each individual photon in
the pair is determined by the density matrix
\begin{equation}\label{reduced}
    \rho_j = \mu\ket{H_j}\bra{H_j}+(1-\mu)\ket{V_j}\bra{V_j},
\end{equation}
where index $j = 1$ and $2$ numbers the subsystems.

In this way, we can prepare a state with the required values of the coefficients in the Schmidt decomposition but in the fixed (H–V) basis. Transformation to an arbitrary basis (\ref{Schmidt}), however, presents no special problems. Actually, an arbitrary polarization state of the qubit, i.e., the polarization state of a photon in a single field mode, can be obtained from a specified state by
means of transformations performed with the use of consequently placed quarter-wave plate (QWP) and half-wave plate (HWP). Let this transformation be $U_j$. Then, we have $\ket{H_j}\xrightarrow{U_j}\ket{A_j}$ and simultaneously $\ket{V_j}\xrightarrow{U_j}e^{i\phi}\ket{B_j}$. The phase $\phi$ can be eliminated by means of plates QP. As a result, we obtain the transformation
\begin{equation}\label{local}
    \ket{\psi}\xrightarrow{U_1\otimes U_2}\sqrt{\mu}\ket{A_1}\ket{A_2}+\sqrt{1-\mu}\ket{B_1}\ket{B_2},
\end{equation}
which ends the procedure of preparing an arbitrary state of a ququart. For experimental convenience, modes $1$ and $2$ are spatially separated due to the non-collinear phase-matching. After the transformations in each spatial–frequency mode are performed, two modes are combined on a dichroic beam splitter (DBS)\footnote{In the dichroic beam splitter, the coefficients of reflection and
transmission are maximum for light beams with the frequencies
$\omega_1$ and $\omega_2$, respectively, regardless of the polarizations.} to obtain a single spatial mode.

The use of frequency non-degenerate regime is an essential feature of this scheme, making it possible to form a collinear beam of biphotons without losses. This is really convenient for practical applications in various quantum information protocols. However, in practice, the preparation of entangled states of ququarts in the scheme with two consequent crystals involves certain problems. These problems are
associated with frequency dispersion effects leading to distinguishability of the photon pairs generated in the first and second crystals and, correspondingly, to the decrease in purity of the prepared state.

\section{Effects of frequency dispersion and their compensation}

Only two frequency modes are singled out in expression (\ref{seed}). Experimentally, this corresponds to the filtration of the initial spectrum with ultimately narrow filters. When using filters with a finite bandwidth, it is necessary to take into account the structure of SPDC frequency spectrum. This leads to a more complex expression for the state vector of the biphoton pair. Let the axes of the first
and second crystals be oriented vertically and horizontally, respectively. Let us consider the case of continuous pumping that is linearly polarized at 45°. In the plane wave pump approximation, the individual states of biphoton pairs generated in each crystal of length $L$ can be written in the form \cite{BurlakovPRA97}
\begin{equation}\label{1crystal-state}
    \ket{\psi_{1,2}}= \int d\Omega d\theta_1 d\theta_2 e^{i\frac{\Delta_z L}{2}} \mathrm{sinc}\left( \frac{\Delta_z(\Omega) L}{2} \right) a^\dag_{V,H}(\theta_1,\omega_1+\Omega) a^\dag_{V,H}(\theta_2,\omega_2-\Omega) \ket{\mathrm{vac}},
\end{equation}
where $\Delta_z(\Omega) =
k_o(\omega_1+\Omega,\theta_1)+k_o(\omega_2-\Omega,\theta_1)+k_e(\omega_p)$
is the longitudinal phase-matching detuning (the pump is an extraordinary wave, and the signal and idler photons are ordinary waves in the crystal). We will analyze the case in which the angular modes are fixed; i.e., they are selected with the use of narrow apertures. Under these conditions, integration over angles is not performed; hence, we can restrict our analysis to the examination of the frequency spectrum. If the pump coherence length exceeds the distance between the crystals \footnote{In practice, this requirement is always satisfied.}, the corresponding amplitudes add coherently and the total state is the superposition
\begin{equation}\label{2crystal-state}
    \ket{\psi}=\frac{1}{\sqrt{2}}\sum_{p=\{H,V\}}\int d\Omega F(\Omega,p)a_p^\dagger(\omega_1+\Omega)a_p^\dagger(\omega_2-\Omega)\ket{\mathrm{vac}},
\end{equation}
where $p$ numbers the polarization states,
\begin{equation}\label{amplitude-1}
F(\Omega,H)\equiv F(\Omega)=\mathrm{sinc}\left( \frac{\Delta_z L}{2}\right)
\end{equation}
and
\begin{equation}\label{amplitude-2}
F(\Omega,V)= F(\Omega)e^{i\varphi(\Omega)}
\end{equation}
are the biphoton amplitudes in the first and second crystals, respectively.

The presence of additional relative phase $\varphi(\Omega)$ is determined by two factors: the phase difference between the polarization components of the pump appearing after the first crystal and the appearance of an additional phase for the photons generated in the first crystal during their passage through the second crystal \cite{ChekhovaPRA07}:
\begin{equation}\label{phase}
    \varphi(\Omega)= (k_e(\omega_p)-k_o(\omega_p))L + (k_e(\omega_1+\Omega)+k_e(\omega_2-\Omega))L.
\end{equation}
The first term on the righthand side of expression (\ref{phase}) does not depend on the phase-matching frequency detuning $\Omega$ and can be easily compensated by introducing the corresponding phase shift between the polarization components of the pump. The second
term is related to the frequency dispersion and determines the appearance of different phase shifts for different frequencies within the spectral linewidth of SPDC. Physically, this means that if a fairly wide part of SPDC spectrum is selected, its individual
components will be added with different phases, which is equivalent to loss of coherence and to generation of a mixed ququart state. The polarization-frequency state will exist only in a Hilbert space of a higher dimension $D \geq 4$. This dimension can be estimated, for example, by using the Fedorov parameter \cite{Fedorov0608} adapted to polarization-frequency distributions; however, this problem is
beyond the scope of this work.

Let us consider the influence of dispersion effects on the properties of generated polarization state. The polarization density matrix, which corresponds to the state $\ket{\psi}$, can be obtained by taking the partial trace over the frequency variables,
\begin{equation}\label{rho}
\rho_{\mathrm{pol}}=\mathrm{Tr}_{\Omega}\ket{\psi}\bra{\psi}=
\frac{1}{2}\sum_{p=\{H,V\}}\int d\Omega F(\Omega,p)
F^*(\Omega,p)\ket{pp}\bra{pp}
\end{equation}
in the obvious notation
$\ket{pp}\equiv
a_p^\dagger(\omega_1+\Omega)a_p^\dagger(\omega_2-\Omega)\ket{\mathrm{vac}}$.
We obtain the following expression for the polarization density matrix:
\begin{equation}\label{r_pol}
    \begin{split}
    \rho_{\mathrm{pol}}=\frac{1}{2} ( \ket{HH}\bra{HH} &+\ket{VV}\bra{VV}\ + \\
    \int d\Omega \left|F(\Omega)\right|^2 e^{-i\varphi(\Omega)}\ket{HH}\bra{VV} &+ \int d\Omega \left|F(\Omega)\right|^2 e^{i\varphi(\Omega)}\ket{VV}\bra{HH}),
    \end{split}
\end{equation}
which, generally speaking, corresponds to a mixed polarization state. The pure state can be obtained only in the case where $\varphi(\Omega)= \mathrm{const}$.

\begin{figure}[h]
\centering\includegraphics[width=0.4\textwidth]{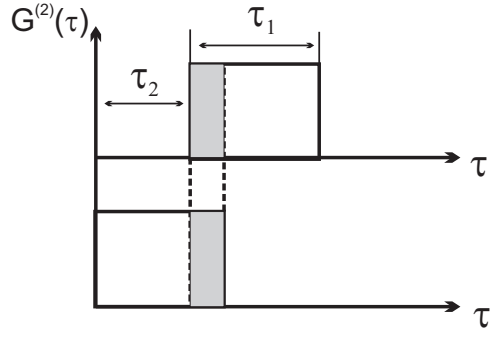}
\caption{{\protect\footnotesize { Second-order correlation function for SPDC from two orthogonally oriented crystals. Regions where correlation functions overlap, leading to interference, are highlighted in grey.}}} \label{g2}
\end{figure}

To the first order in frequency detuning $\Omega$, we have
\begin{equation}\label{Delta}
    \Delta_z(\Omega)=\left[ \left(\frac{\partial k_o(\omega)}{\partial \omega}\right)_{\omega_1} - \left(\frac{\partial k_o(\omega)}{\partial \omega}\right)_{\omega_2}\right]\Omega = \left[ \frac{1}{v^{gr}_o(\omega_1)} - \frac{1}{v^{gr}_o(\omega_2)} \right]\Omega = C_o\Omega,
\end{equation}
\begin{equation}\label{Phi}
    \varphi(\Omega) = \left[ \left(\frac{\partial k_e(\omega)}{\partial \omega}\right)_{\omega_1} - \left(\frac{\partial k_e(\omega)}{\partial \omega}\right)_{\omega_2}\right]L\Omega = \left[ \frac{1}{v^{gr}_e(\omega_1)} - \frac{1}{v^{gr}_e(\omega_2)} \right]L\Omega = C_eL\Omega.
\end{equation}
Here,  $v^{gr}_{o,e}(\omega)=\left(\frac{\partial\omega}{\partial
k_{o,e}}\right)$ is the group velocity of wave packets with the corresponding central frequencies. Expression (\ref{Delta}) determines the shape of SPDC spectral line: $\left|F(\Omega)\right|^2=\mathrm{sinc}^2\left(C_oL\Omega/2\right)$.
In our case, it is more convenient to consider the temporal representation and to turn from spectral characteristics to temporal correlations. Second-order correlation function and spectrum are related by the following expression \cite{ChekhovaJETP02}:
\begin{equation}\label{CorrF}
    G^{(2)}(\tau)=\left| \int d\Omega F(\Omega)\cos(\Omega\tau) \right|^2.
\end{equation}
When the function $F(\Omega)$ is symmetric, we obtain a conventional Fourier transform. Therefore, in our
case, the correlation function of biphotons from each crystal has the form of a rectangle with a width of
$\tau_1 = C_o L$. The presence of the phase shift $\varphi(\Omega)$ in temporal terms implies a relative shift of the correlation functions by $\tau_2 = C_e L$ (see Fig. \ref{g2}).

\begin{figure}[h]
\centering\includegraphics[width=0.8\textwidth]{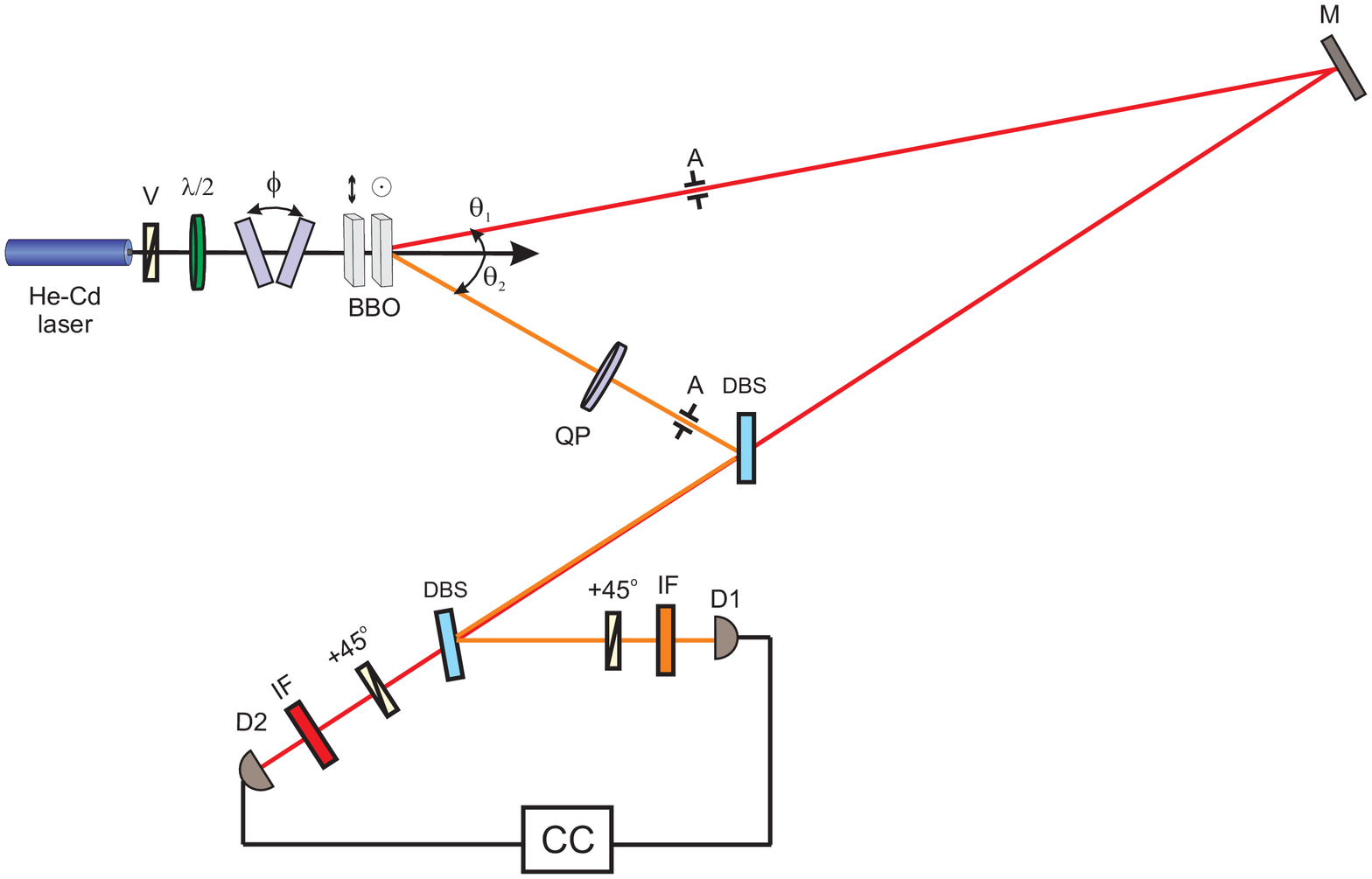}
\caption{{\protect\footnotesize { Experimental setup.}}}
\label{setup-bell}
\end{figure}

The relative shift in the correlation functions of biphotons from the first and second crystals leads to their temporal distinguishability, and hence, to a decrease in visibility of polarization interference. Let us consider this problem in detail from the
experimental point of view. Scheme of the experimental setup is depicted in Fig. \ref{setup-bell}.
Helium–cadmium laser with a wavelength of 325 nm was used as a pump. Vertical polarization was selected from the initially unpolarized radiation with the use of a Glan prism (V), after which it was rotated with a half-wave plate. Relative phase $\phi$ between the polarization components of the pump radiation was introduced by two quartz plates with vertically oriented optical axes. The effective thickness of these plates was determined by the relative tilt angle $\theta$. The parametric down-conversion occurred in two BBO crystals (2 mm thick each) cut for the collinear frequency-degenerate phase-matching. The directions corresponding to the operating wavelengths $\lambda_1=600$ nm
and $\lambda_2=710$ nm were separated by the apertures (A). Then, the beams were combined on a dichroic beam splitter (DBS), transmitting radiation at a wavelength of 600 nm and completely reflecting at 710 nm. In the measurement part of the scheme, the pairs were separated in frequency with the use of a similar beam splitter. Film polarizers oriented at 45° with respect to the vertical and interference filters with central wavelengths of 600 nm and 710 nm and a bandwidth of 10 nm (IF) were inserted in each channel of the measurement scheme. The coincidences of photocounts of detectors D1 and D2 were recorded using a coincidence circuit (CC) with a window of 2 ns.

In ideal case, when the polarization state of a
biphoton has the form
\begin{equation}\label{psi_1}
\ket{\psi}=1/\sqrt{2}\left( \ket{H_1H_2}+e^{i\phi}\ket{V_1V_2}
\right),
\end{equation}
the coincidence counting rate in this scheme is given
by the expression
\begin{equation}\label{R_C}
R_C\propto\left|
\bra{+45_1,+45_2}\left({\ket{H_1H_2}+e^{i\phi}\ket{V_1V_2}}\right)
\right|^2\propto1+\cos{\phi},
\end{equation}
where the phase $\phi$ changes with variation in the tilt of the quartz plates in the pump beam. In practice,
since the pairs generated in the first and second crystals are partially distinguishable due to the incomplete
overlap of the corresponding correlation functions (see Fig. \ref{g2}), the visibility of the interference pattern,
which is defined as
\begin{equation}\label{Vis1}
V=\frac{{R_C}_{max}-{R_C}_{min}}{{R_C}_{max}+{R_C}_{min}},
\end{equation}
differs from unity substantially. By using (\ref{r_pol}) for the polarization density matrix, we can derive the following expression for the coincidence counting rate:
\begin{equation}\label{R_1}
    \begin{split}
    R(\phi)\propto 1+1\int d\Omega \left|F(\Omega)\right|^2 \cos(\varphi(\Omega)-\phi) = \\
    1+V\cos(\phi),
    \end{split}
\end{equation}
where the visibility of the interference pattern (to the
first order in detuning) may be written in the form
\begin{equation}\label{Vis2}
V=\int d\Omega \mathrm{sinc}^2
(\frac{C_oL\Omega}{2})\cos(C_eL\Omega).
\end{equation}
The integral is calculated simply and leads to the following expression for visibility:
\begin{equation}\label{Vis3}
V=1-\frac{C_e}{C_o}=1-\frac{\tau_2}{\tau_1},
\end{equation}
which is consistent with the intuitive expectations illustrated in Fig. \ref{g2}.

For a fixed frequency spectrum, which, in our case, is determined by the bandwidth of the interference filters (it appears to be somewhat smaller than the width of the SPDC spectrum, which, in the regime used, is approximately 12 nm), the visibility can be increased
by introducing an additional birefringent compensator QP into the beam with a shorter wavelength. Quartz compensator with vertically oriented optical axis introduces a delay for the vertically polarized photon which is represented by the expression
\begin{equation}\label{tau1}
\tau_{comp}=\left[ \right(\frac{1}{{v^{gr}_e}_q (\omega_1)} -
\frac{1}{{v^{gr}_o}_q (\omega_1)}\left)L_q \right],
\end{equation}
where ${v^{gr}_{o(e)}}_q$ are the group velocities of the ordinary and extraordinary waves in quartz, respectively, and $L_q$ is the compensator length. By choosing the compensator length to satisfy the condition $\tau_{comp}=\tau_2$, it is possible to achieve complete overlap of the correlation functions and, consequently, maximum visibility of the interference pattern.

\begin{figure}[h]
\centering\includegraphics[width=0.8\textwidth]{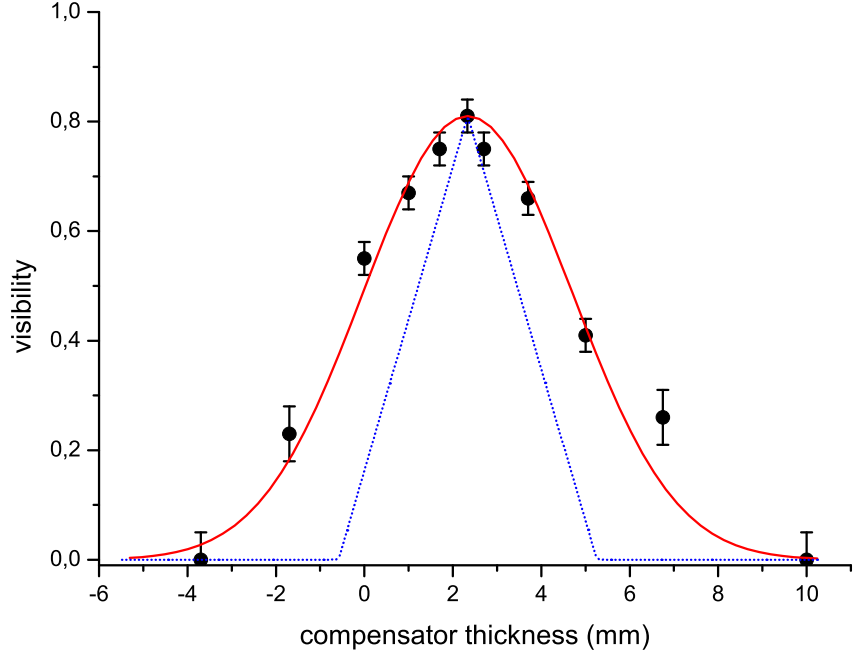}
\caption{{\protect\footnotesize {
Dependence of polarization interference visibility on thickness of group velocity dispersion compensator. Dashed curve is the theoretical dependence not taking filtering into account. Solid curve is numerically calculated dependence taking 10 nm bandwidth interference filters in each channel into account.}}}
\label{visibility}
\end{figure}

Figure {\ref{visibility}} shows the experimental dependence of the visibility of interference pattern on the length of
the quartz compensator used. The dotted curve bounding the triangle corresponds to the convolution of two rectangular correlation functions, i.e., the functions determined by the shape of the SPDC spectral line only, without filtration. However, the experiment was performed using interference filters characterized by a Gaussian shape of the dependence of transmission coefficient on the wavelength with the width comparable to that of the SPDC spectrum. Therefore the correlation functions should have a shape somewhat different from rectangular. When the frequency filtration is taken into account, the expression for the biphoton amplitude additionally acquires the factor corresponding to the transmission amplitude of the interference filters. By choosing this factor to be Gaussian, we obtain
\begin{equation}\label{ampl3}
F(\Omega)\propto
\exp(-\frac{(\omega_1-\Omega)^2}{\Delta\omega^2}) \exp(-\frac{(\omega_2-\Omega)^2}{\Delta\omega^2}) \mathrm{sinc}\left(
\frac{\Delta_z(\Omega) L}{2} \right),
\end{equation}
where $\Delta\omega$ is the interference filters bandwidth. The solid curve in Fig. \ref{visibility} represents the numerically
calculated dependence of the visibility on the group velocity dispersion compensator thickness taking the 10 nm interference
filters into account.

\begin{figure}[h]
\centering\includegraphics[width=\textwidth]{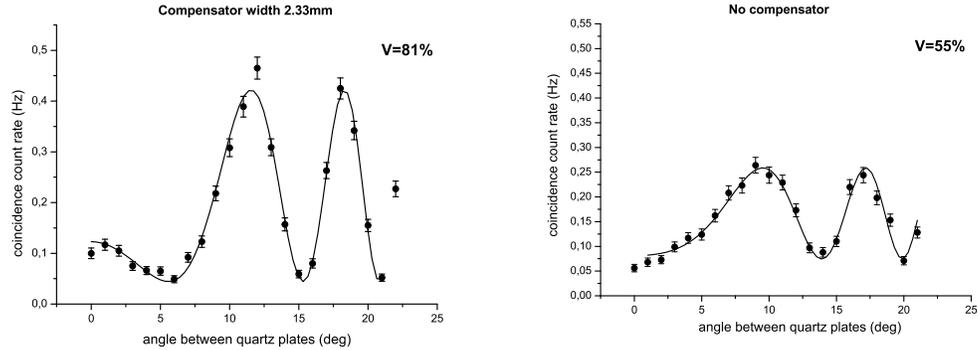}
\caption{{\protect\footnotesize {Interference patterns obtained without compensation and with a 2.33 mm thick compensator.}}}
\label{interf}
\end{figure}

The calculation of width and shift of the correlation functions in the case under consideration leads to the following values: $\tau_1=96$ fs, $\tau_2=87$ fs, which corresponds to a compensator thickness of 2.6 mm, in good agreement with experimental data. The small difference between the experimental position of the maximum in the visibility (2.3 mm) and the calculated
value is most likely explained by the errors in the values of group velocities used. The characteristic interference patterns obtained with and without a compensator are shown in Fig. \ref{interf}.

The the non-absolute visibility even with the use of a compensator of required thickness is explained by two factors. The first factor is "technical" and lies in the inaccuracy of selecting the necessary spatial modes. The second factor is the influence of the
terms quadratic in the frequency detuning $\varphi(\Omega)$ in the expansion of $\varphi(\Omega)$, which lead to unequal broadenings of the correlation functions and, consequently, to the deviation of visibility from 100\% even in the case of ideal compensation for the part that is linear in the detuning \cite{ChekhovaPRA07}. Indeed, to the second order in the detuning from the degenerate frequency, we have
\begin{equation}\label{Phi_2}
    \varphi(\Omega)=\frac{1}{2}\left(\left(\frac{\partial^2 k_e}{\partial\omega^2}\right)_{\omega_1}+ \left(\frac{\partial^2 k_e}{\partial\omega^2}\right)_{\omega_2}\right)L\Omega^2 = B_eL\Omega^2.
\end{equation}
If we denote $\xi=C_oL\Omega$ и $D=\frac{B_e}{C_o^2L}$ than for the coincidence counting rate in the case
of a complete compensation of the first-order effects, we have
\begin{equation}\label{R_2}
    R(\phi)\propto 1+\frac{1}{2\pi}\left(V_c\cos\phi+V_s\sin\phi\right),
\end{equation}
where $V_c=\int d\xi \mathrm{sinc}^2 \frac{\xi}{2}\cos D\xi^2$ и $V_s=\int d\xi\mathrm{sinc}^2\frac{\xi}{2}\sin D\xi^2$.
For the crystals used in the experiment, we have $D = 0.0193$ and numerical calculation gives the following value of interference pattern visibility:
\begin{equation}\label{Vis4}
V=\frac{\sqrt{V_c^2+V_s^2}}{2\pi}=0.89.
\end{equation}

\begin{figure}[h]
\centering\includegraphics[width=0.5\textwidth]{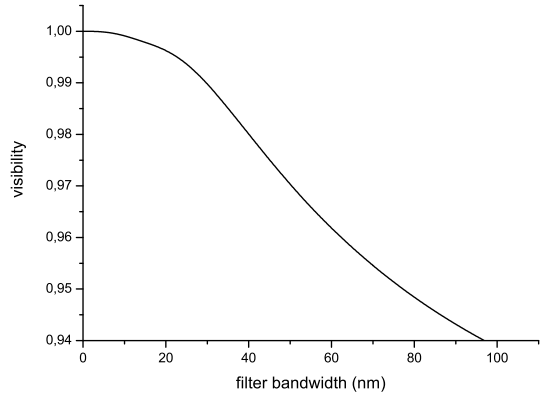}
\caption{{\protect\footnotesize { Dependence of polarization interference visibility on the interference filter bandwidth in the case of complete group delay compensation.}}}
\label{visibility_2}
\end{figure}

It should be noted that such a low value \footnote{The record-high values of the visibility in experiments with narrowband filters in the frequency degenerate regime reach 99\%.} of visibility is the lower limit, corresponding to complete absence of frequency filtration. The results of numerical calculation of the dependence of visibility on the bandwidth of interference filters used in the case of a complete compensation are presented in Fig. \ref{visibility_2}. It can be seen from this figure that, even when sufficiently broadband interference filters are used (with a bandwidth of less than 30 nm), the values of visibility can reach 0.95 or higher.

The low visibility of polarization interference corresponds to poor quality of entangled states preparation. The currently obtained results of statistical state reconstruction for the state
\begin{equation}\label{phi+}
\ket{\Phi^+}=1/\sqrt{2}\left( \ket{H_1H_2} \\ +\ket{V_1,V_2}
\right),
\end{equation}
which corresponds to the maximum of the interference pattern shown in Fig. \ref{interf}, lead to a fidelity, defined as
\begin{equation}\label{fidelity}
F = | \langle \psi_{theory}| \psi_{exp}\rangle|^2,
\end{equation}
not exceeding 75\%. Nonetheless, we do not see, in principle, any restrictions from above on this quantity. Higher values of fidelity can be reached by means of careful selection of spatial modes and, possibly, with the use of narrower band frequency filters. However, the mere fact that we observe the two-photon interference with a nonzero fidelity suggests that the preparation of entangled states according to the proposed scheme is realistic. Regarding improvement of the prepared states quality, it is not a crucial and rather technical problem, wich we are working at now.

\section{Conclusions}

We have considered a method for preparing arbitrary entangled states of polarization ququarts in non-collinear frequency non-degenerate type-I SPDC regime according to the scheme with two consequent crystals. We have elucidated the role of effects that are related to the frequency dispersion and which lead to a decrease in purity of the prepared states and, as a consequence, to a decrease in visibility of the two-photon polarization interference. It has been shown that, in the frequency non-degenerate regime, the compensation of group velocity dispersion, which was previously used in schemes with pulse pumping only, is necessary even in the case of continuous pump with large coherence length. Methods providing this compensation have been proposed and demonstrated experimentally. Moreover, in the case of a complete compensation, the maximum value of visibility limited by dispersion effects of higher orders in the frequency
detuning has been estimated.

We would like to emphasize that the application of frequency non-degenerate biphotons, unlike the majority of earlier proposed and currently known schemes (using frequency degenerate biphotons and spatial modes selection), offers a number of advantages.
The main advantage is the possibility of forming states, including entangled ones, which are localized in a single spatial mode. This property is definitely desirable from the viewpoint of possible practical applications, when two-particle states must be transmitted through a single communication channel. Thus, current work has revealed physical restrictions on the quality of prepared states and demonstrated methods for its improvement using the particular experimental scheme.

We would like to thank D. Ivanov for checking the
calculations.

This work was supported in part by the Russian Foundation for Basic Research (project nos. 08-02-00741-a and 08-02-00559-a), NATO (grant no. CBP.NR.NRCL 983251), and the Council on Grants from the President of the Russian Federation for the Support of Leading Scientific Schools (grant no. NSh-796.2008.2).

\end{document}